\documentstyle[aps]{revtex}
\begin{document}
\draft
\title{GEOMETRIZATION OF VACUUM CONDENSATE EFFECTS IN QUARKONIUM POTENTIAL MODEL
\thanks{Work presented at PANIC99 conference, Uppsala, Sweden (June
10 -- 16, 1999) }}
\author{A.V. Shoulgin
\thanks{ashoulgi@uic.rnd.runnet.ru}
\\ }
\address{Department of Physics, Rostov State University, Zorge
5, Rostov-on-Don, 344090, RUSSIA}
\maketitle
\begin{abstract}
It is suggested the modification of traditional potential model,
in which nontrivial structure of inside-hadron vacuum condensate
is simulated by geometric properties of inside-hadron space.
Confinement of quarks is ensured by closed effective (Riemannian)
space. Interquark potential represents itself Coulomb's low in the
effective space (Poisson equation). In the framework of such
approach the quarks dynamics completely is defined only by metric
of the effective space, which in turn in conformally-Euclidean
case (we consider) is defined by sole phenomenological function.
Also it is supposed, that inside-hadron vacuum is essential
nonperturbative one both on large and on small distances and this
is taken into account by special parameter in the metric of
effective space. For final Schr\"odinger equation the exact
analytical solutions for nonrelativistic energy spectrum and wave
functions of quarkonium are obtained. From the basic principles of
geometrized potential model and under the perturbation theory the
spin-dependent relativistic corrections are calculated. Charmonium
and bottomonium spectra are simulated. Suggested model gives a
very good fit to experimental data: accuracy of spectra fitting
makes $4.06\cdot 10^{-2}$ for charmonium and $3.28\cdot 10^{-2}$
for bottomonium and many predictions are made. On the basis of
charmonium and bottomonium analysis conclusions about role of
vacuum condensate in hadron structure are done.
\end{abstract}
\section{Introduction.}
Quarkonia are bound states of heavy quarks and antiquarks $c\bar
c$, $b\bar b $, $t\bar t$ form the special class of objects in the
hadrons physics. Experimentally observable quarkonia properties
are used for clearing up of static and dynamic quark properties.
The quarkonia spectroscopy is studied in detail experimentally
\cite{data} and gives in to effective theoretical modeling. The
nonrelativistic character of heavy quarks moving within quarkonia
allows to apply for calculation of their spectra nonrelativistic
Schr\"odinger equation, in which effective potential of quarks
interaction $U(r)$ and quarks reduced mass $\mu (r)$ as functions
of coordinates are selected phenomenologically. Criterion of
functions $U(r)$ and $\mu(r)$ choice is a good fit of potential
model outcomes with experimental data. Smallness of relativistic
effects allows to specialize nonrelativistic potential models by
inserting to Hamiltonian effects of spin-orbit and spin-spin
interactions as small corrections computed under perturbations
theory. Modern condition of potential models is reflected in the
review paper \cite{Besson}. For more detailed information see
lectures \cite{Lucha}.

Central object in realization of the nonrelativistic potential
program is the potential of interquark interaction $U(r)$, which
should take into account effect of absolute quarks confinement
within hadron. Its choice is ambiguous in view of unsufficient
methods of the interquark interactions analysis on large
distances. The potential is selected usually from some physical
reasons in combination with the aesthetic requirements of
simplicity. The existing theoretical indications concerning the
form of potential most explicitly are reflected in so-called
Cornell potential \cite {cornel}
\begin{equation}
\label{Cornell}U(r)=-\frac{g^2}r+\sigma r,
\end{equation}
where the first term is motivated by perturbative QCD with $r\to
0$ with an accuracy to logarithmic corrections, taking into
account the perturbative vacuum polarization effects; second term
is predicted by nonperturbative lattice QCD with $r\to \infty $ .
Majority of used potentials differ among themselves by character
of interpolation on intermediate distances, between two limiting
conditions reflected in the formula (\ref{Cornell}).

Concerning of reduced mass, as function of coordinates, only its
asymptotic on small distances, found in perturbative QCD, is
known. For potential model, however the dependence of reduced
masses from coordinates on average and large distances is
necessary, which unfortunately theoretically is not found. For
this reason the reduced mass is simulated, as a rule, by constant
parameter, which value is determined from the requirement of best
agreement of theoretical outcomes with experiment.

The use of the above-stated procedures allows in whole to achieve
a good quantitative agreement of the theory with experiment,
however, a problem of physical completeness and justification of
the potential approach remains open. In the framework of the
fundamental QCD hadron is an area of reconstructed vacuum -- is a
cavity inside nonperturbative quark-gluon condensate stabilized by
valent quarks, residing within it. Moreover, the inside-hadron
reconstructed vacuum exert influence on quarks dynamics, exactly
on reduced mass and interquark potential. Thus, quarkonium is
complicated self-consistent object, in which quarks condition
depends from condition of reconstructed nonperturbative vacuum and
condition of vacuum inside quarkonium depends from quarks
condition. In this situation it is worth to suggest the new
phenomenological approaches in the theory of quarkonia. The more
general potential models, at first should contain additional gang
of functional and numerical parameters in the interquark
potential, supposing the interpretation in the terms of
nonperturbative vacuum physics; secondly, in these models the
problem of reduced mass of quarks within nonperturbative
inside-hadron vacuum and problem of interquark potential should be
considered from uniform positions.

In present work one of the variants of new potential models,
possessing by the above-stated properties, is suggested. Our
approach is based on the consideration of inside-hadron space as
manifold with the closed topology. The quarks reduced mass and the
interquark potential completely are defined by geometric
properties of the effective Riemannian space; metric of this space
represents itself as universal phenomenological functional
parameter. We also assume, that vacuum inside quarkonium is
essential nonperturbative one both on large and on small distances
and take this into account by specially selected parameter in the
metric of effective space, which then occurs in quarks reduced
mass and interquark potential functional dependencies from
coordinates of real (Euclidean) space.

The paper is constructed as follows. In section II the basic
notions about vacuum element in hadron are stated; in section III
the geometrization of potential model is carried out and its
mathematical structure is stated; in section IV the generalized
Cornell potential, parametric taking into account nonperturbative
structure of vacuum on all distances is introduced; here the exact
analytical solutions for nonrelativistic Schr\"odinger equation
are obtained; in section V the relativistic effects are
considered; section VI is devoted to the results and conclusions.

\section{The vacuum element in the hadron mass.}

In the framework of the qualitative script of the hadron formation
the reconstruction of vacuum condensate, described quite
determined energy, leading to vacuum constituting element in
hadron mass, as a necessary
condition is contained. This vacuum constituting element reads%
$$ E_{vac}=\int \varepsilon _{vac}(\vec r)dV $$ where $$
\varepsilon _{vac}(\vec r)=\varepsilon _{in}(\vec r)-\varepsilon
_{out}>0, $$ $\varepsilon _{out}$ is the energy density of
outside-hadron vacuum condensate (nonperturbative quark-gluon
condensate), $\varepsilon _{in}(\vec r)$ is the energy density of
inside-hadron (reconstructed) vacuum condensate, here the
coordinate dependence takes into account the spatial heterogeneity
of condensate. The mass of hadron is represented as follows:
\begin{equation}
\label{m}M=E_{vac}+E_q,
\end{equation}
where $E_q$ is a sum of rest energy, kinetic energy and energy of quarks
interactions with each other and with vacuum condensate. The birth threshold
of hadron agrees (\ref{m}) is defined not only by quark masses, but also
energy $E_{vac}$, expended to quite defined reconstruction, after which
inside a new vacuum the processes of quark excitations births with masses $%
m_q$ become possible.

\section{The model.}

Let us to start from quarkonium Lagrangian, which in the framework of
potential approach reads
\begin{equation}
\label{1}L=\frac 12\mu (\vec r)\left( \frac{dl_E}{dt}\right) ^2-U(\vec
r),\quad dl_E^2=dr^2+r^2(d\theta ^2+\sin ^2\theta d\varphi ^2),
\end{equation}
where $\mu (\vec r)$ is the quarks reduced mass and $U(\vec r)$ is
the quark-antiquark potential.

Having limited by a class of isotropic phenomenological functions $\mu
(r)=mf^2(r)/2$,$\;U(r)$, let's rewrite Lagrangian (\ref{1}) as
\begin{equation}
\label{2}L=\frac m4\left( \frac{dl}{dt}\right) ^2-U(r),
\end{equation}
\begin{equation}
\label{3}dl^2=f^2(r)(dr^2+r^2(d\theta ^2+\sin ^2\theta d\varphi ^2))\equiv
\gamma _{\alpha \beta }dx^\alpha dx^\beta .
\end{equation}
The first term in (\ref{2}) formal coincides with kinetic energy of particle
with mass $m/2=const$, moving in curved Riemannian space with metric (\ref{3}%
). It means the condensate influence to the quark inert properties
can be mathematically taken into account by transition from real
Euclidean space to effective Riemannian space. Function $U(r)$,
appearing in (\ref{2}), within the framework of accepted
hypothesis is identified with Coulomb's law in Riemannian space
with metric (\ref{3}):
\begin{equation}
\label{Cou}\frac 1{\sqrt{\gamma }}\frac \partial {\partial x^\beta }\sqrt{%
\gamma }\gamma ^{\alpha \beta }\frac{\partial U}{\partial x^\beta }\equiv
\frac 1{r^2f^3(r)}\frac d{dr}r^2f(r)\frac{dU}{dr}=g^2\delta (\vec r).
\end{equation}
From the equation (\ref{Cou}) follows
\begin{equation}
\label{F}\mu (r)=\frac m2\left( \frac{r^2}{g^2}\frac{dU}{dr}\right)
^{-2},\qquad U(r)=g^2\int\limits_{r_0}^r\frac{dr}{r^2f(r)}.
\end{equation}
The effective potential in (\ref{F}) is defined with an accuracy
to additive constant, which value is parameterized by a limit
inferior in quadrature. Such model's ambiguity, however, is
insignificant, since this additive constant is swallowed by vacuum
energy (\ref{m}), assigning count beginning of energy spectrum.

For geometrized potential model the hypothesis about additional
quark-condensate interaction, which energy proportional to
curvature of the effective Riemannian space is mathematically
natural. The insertion this interaction to model lead up to
Lagrangian
\begin{equation}
\label{lag}L=\frac m4\gamma _{\alpha \beta }\frac{dx^\alpha }{dt}\frac{%
dx^\beta }{dt}-U(r)+\kappa R(r),
\end{equation}
where $\kappa $ - phenomenological quark-condensate coupling
constant;
\begin{equation}
\label{curve}R(r)=2\left[ \frac 1{f^4}\left( \frac{df}{dr}\right) ^2-\frac
2{f^3}\frac{d^2f}{dr^2}-\frac 4{rf^3}\frac{df}{dr}\right]
\end{equation}
is scalar curvature, evaluated in terms of coordinates (\ref{3}).

For Lagrangian (\ref{lag}) the corresponding Schr\"odinger
equation is
\begin{equation}
\label{Shr}\left( -\frac 1m\gamma ^{\alpha \beta }\nabla _\alpha \nabla
_\beta +U(r)-\kappa R(r)\right) \Psi =E\Psi
\end{equation}
where $\nabla _\alpha $ is covariant derivation operator in space with
metric (\ref{3}). After variable separation
\begin{equation}
\label{factorization}H\Psi _{nLm}=E_{nL}\Psi _{nLm}\;,\;\Psi _{nLm}(r,\theta
,\phi )=R_{nL}(r)Y_{Lm}(\theta ,\phi ),
\end{equation}
equation (\ref{Shr}) is reduced to equation for the radial part of
wave function
\begin{equation}
\label{Shrodinger}-\frac 1m\frac 1{r^2f^3}\frac d{dr}r^2f\frac{dR_{nL}}{dr}+\left[\frac 1m\frac{%
L(L+1)}{r^2f^2}+ g^2\int\limits_{r_0}^r\frac{dr}{r^2f}-2\kappa
\left( \frac 1{f^4}\left( \frac{df}{dr}\right) ^2-\frac 2{f^3}\frac{d^2f}{%
dr^2}-\frac 4{rf^3}\frac{df}{dr}\right) \right] R_{nL}=E_{nL}R_{nL}
\end{equation}
with invariant normalization
\begin{equation}
\label{Norma}\int \mid \Psi _{nLm}\mid ^2\sqrt{\gamma }d^3x=\int\limits_0^%
\infty R_{nL}^2(r)f^3(r)r^2dr=1.
\end{equation}

From the view point of physical representations, explained above,
the choice of non-Euclidean measure of integration in
(\ref{Norma}) is finishing mathematical operation under the
account of condensate influence on spatial localization of
quarkonium state.

The radial coordinate $r$, used in (\ref{2}) -- (\ref{Norma}), is
common both for the real Euclidean space and for the effective
Riemannian space. Therefore, arguing of problem about radial
dependence of quark-antiquark potential is carried out in terms of
this coordinate. However, in mathematical research of the equation
(\ref{Shrodinger}) it is more convenient to use dimensionless
coordinate $\rho $ and phenomenological function $\chi (\rho )$:
\begin{equation}
\label{ro}\rho =\frac 1a\int\limits_{r_0}^rf(r)dr\;,\;\chi (\rho )=\frac
1ar(\rho )f\left[ r(\rho )\right] \;,\;\rho _{max}=\rho (\infty )
\end{equation}
The metric (\ref{3}), curvature (\ref{curve}) and Coulomb's law (\ref{F})
become:
$$
dl^2=a^2\left[ d\rho ^2+\chi ^2(\rho )\left( d\theta ^2+\sin ^2\theta d\phi
^2\right) \right] \equiv \gamma _{\alpha \beta }dx^\alpha dx^\beta
$$
\begin{equation}
\label{RU}R(\rho )=\frac 2{a^2}\left[ \frac 1{\chi ^2}-2\frac{\chi ^{\prime
\prime }}\chi -\left( \frac{\chi ^{\prime }}\chi \right) ^2\right]
,\;\;U(\rho )=g^2\int\limits_{\rho _0}^\rho \frac{d\rho }{\chi ^2(\rho )}
\end{equation}
In (\ref{RU}) and in the further primes derivative are designated on $\rho$.
The physical and the geometrical meaning of dimensional parameter $a$ become
clear after giving representations for potential, satisfying to condition of
quarks confinement within hadron.

Further, it is convenient to turn to new radial function with simple
normalization:
\begin{equation}
\label{normalization}R_{nL}=\frac 1{a^{3/2}\chi (\rho )}\phi _{nL}(\rho
)\;,\;\int\limits_0^{\rho _{max}}\phi _{nL}^2(\rho )d\rho =1
\end{equation}
and to dimensionless quantities:
\begin{equation}
\label{less}\varepsilon _{nL}=\frac{ma^2}{\hbar ^2}E_{nL},\;\;2\gamma =\frac{%
g^2ma}{\hbar ^2},\;\;\lambda =\frac{2\kappa m}{\hbar ^2}.
\end{equation}
Now equation (\ref{Shrodinger}) corresponds as follows:
\begin{equation}
\label{itog}\phi _{nL}^{\prime \prime }+\left[ \varepsilon _{nL}-2\gamma
\int\limits_{\rho _0}^\rho \frac{d\rho }{\chi ^2}-\frac{L(L+1)}{\chi ^2}-%
\frac{\chi ^{\prime \prime }}\chi +\lambda \left( \frac 1{\chi ^2}-2\frac{%
\chi ^{\prime \prime }}\chi -\frac{\chi ^{^{\prime }2}}{\chi ^2}\right)
\right] \phi _{nL}=0
\end{equation}

\section{The generalized Cornell potential and exact solutions of the
Schr\"odinger equation.}

In quarkonium geometrized potential model the radial dependence of
the effective potential is set simultaneously with the metric and
the curvature of effective Riemannian space by sole
phenomenological function:
\begin{equation}
\label{uniq}f(r)=\left( \frac{r^2}{g^2}\frac{dU}{dr}\right) ^{-1}.
\end{equation}
Among of potentials satisfying to condition of quarks confinement
within hadron the potentials of whirlpool form stands out of the
simplicity:
\begin{equation}
\label{potential}U(r)=\frac{g^2}{2a}\left[ -\left( \frac{2a}r\right)
^{1/k}+\left( \frac r{2a}\right) ^{1/k}\right] ,
\end{equation}
Where $2a$ is characteristic quarkonium size (''diameter''); $k$
is positively defined numerical parameter. With $k=1$ expression
(\ref{potential}) passes to Cornell potential \cite{cornel},
joining known asymptotics. Euclidean-Coulomb asymptotic of Cornell
potential with $r\to 0$ corresponds to supposition that properties
of inside-hadron vacuum in a neighborhood of point $r=0$ limiting
are closed to properties of perturbative vacuum. It is possible to
expect, with $k\ne 1$ generalized Cornell potential (\ref
{potential}) simulates situation with nonperturbative
inside-hadron vacuum condensate. For potential (\ref{potential})
the evaluations under the formulas (\ref{uniq}), (\ref{ro}),
(\ref{RU}) give: $$ f(r)=\frac{k}{\left[ \left( \frac{2a}r\right)
^{\frac 1k-1}+\left( \frac r{2a}\right) ^{\frac 1k+1}\right]},
\quad r=2a\tan ^k\frac \rho {2k^2},\quad 0\le \rho <{\pi }k^2 $$
\begin{equation}
\label{funcs}\chi (\rho )=k\sin \frac \rho {k^2},\quad U(\rho )=-\frac{g^2}%
a\cot \frac \rho {k^2},\quad R(\rho )=\frac 2{k^4a^2}\left( 3-\frac{1-k^2}{%
\sin ^2\frac \rho {k^2}}\right) .
\end{equation}
The equation (\ref{itog}) and normalization (\ref{normalization})
become:
\begin{equation}
\label{last}\frac{d^2\phi _{nL}}{d\eta ^2}+\left\{ k^4\varepsilon
_{nL}+3\lambda +1+2\gamma k^4\cot \eta -\frac 1{{\sin }^2\eta }\left[
k^2L(L+1)+\nu (\nu +1)\right] \right\} \phi _{nL}=0,
\end{equation}
\begin{equation}
\label{normirovka}k^2\int\limits_0^\pi \phi _{nL}^2(\eta )d\eta =1,
\end{equation}
where
$$
\eta =\rho /k^2,\quad \nu (\nu +1)=\lambda (1-k^2).
$$
The solutions of the equation (\ref{last}) satisfy to condition of absolute
quarks confinement within hadron, if:
\begin{equation}
\label{condition}\lambda (1-k^2)\nu =-\frac 12+\sqrt{\frac 14+\lambda (1-k^2)%
}\ge 0.
\end{equation}
The inequality (\ref{condition}) is the sole restriction on the
phenomenological parameters of geometrized model appropriating to
generalized Cornell potential (\ref{potential}). With $k=1$ the
effective geometry, as it is visible from (\ref{funcs}),
represents a closed homogeneous and isotropic space with constant
positive curvature $R=6/a^2$. The equation (\ref{last}) in this
case describes hydrogen-like system in such space:
\begin{equation}
\label{hydro}\phi _{nL}^{\prime \prime }+\left[ \varepsilon _{nL}+3\lambda
+1+2\gamma \cot \rho -\frac{L(L+1)}{\sin ^2\rho }\right] \phi _{nL}=0.
\end{equation}
From symmetry reasons is beforehand obvious, that spectrum of such
system depends only from the sum of radial and orbital numbers
$N=n+L$, i.e. degenerates on orbital number.

With $k\ne 1$ the degeneracy is removed, but mathematical structure of the
equation for wave function on comparison with (\ref{hydro}) practically does
not vary. Really, after insertion variable $\eta $ and intermediate
auxiliary parameter $l$:
$$
l(l+1)=\nu (\nu +1)+k^2L(L+1)\;,\;l=-\frac 12+\sqrt{\left( \nu +\frac
12\right) ^2+k^2L(L+1)},
$$
the equation (\ref{last}) corresponds as follows
\begin{equation}
\label{verylast}\frac{d^2\phi _{nL}}{d\eta ^2}+\left[ k^4\varepsilon
_{nL}+3\lambda +1+a\gamma k^4\cot \eta -\frac{l(l+1)}{\sin ^2\eta }\right]
\phi _{nL}=0,
\end{equation}
and differs from (\ref{hydro}) only by overdetermination of dimensionless
quantities $\gamma \to \gamma k^4,\;\varepsilon _{nL}\to \varepsilon
_{nL}k^4\;,\;L\to l$. The mathematical uniformity of the equations (\ref
{hydro}) and (\ref{verylast}) does not cancel, however, their physical
distinctions: in (\ref{verylast}) $l$ is not orbital quantum number,
accepting only integer value. In model with potential (\ref{potential}) and
with additional quark-condensate interaction, proportional to curvature,
with $k\ne 1$, parameter $l$ with realization of the condition (\ref
{condition}) can accept any positive values even for states with real
orbital moment $L=0$. The solutions of the equation (\ref{verylast}) are
energy spectrum:
\begin{equation}
\label{solut}\varepsilon _{nL}=-\frac 1{k^4}(1+3\lambda )+\frac
1{k^4}(n+l)^2-\frac{\gamma ^2k^4}{(n+l)^2},
\end{equation}
and exact eigenfunctions
\begin{equation}
\label{exact}\phi _{nL}=\left[ \prod_{s=1}^n\left( -\frac \partial {\partial
\eta }+\frac{(l+s)\cos \eta }{\sin \eta }-\frac{\gamma k^4}{l+s}\right)
\right] \varphi _{nL}.
\end{equation}
In (\ref{exact}) line-ups of operators is ordered in such a way, that
operators, corresponding to smaller values $s$, stand to the left of
operators, corresponding to large $s$. The auxiliary functions $\varphi
_{nL} $ satisfy to equations
$$
\left( -\frac \partial {\partial \eta }-\frac{(l+n)\cos \eta }{\sin \eta }+%
\frac{\gamma k^4}{l+n}\right) \varphi _{nL}=0, $$ having exact
solutions, $$ \varphi _{nL}=C_{nL}\exp (-\frac{\gamma \eta
}{l+n})\sin ^{n+l}\eta , $$
where $C_{nL}$ - normalizing
constants, computed numerically from normalization
(\ref{normirovka}).

Turning to dimensional quantities under the formulas (\ref{less}) and
choosing additive constant in the correspondence with (\ref{m}), we shall
receive from (\ref{solut}) the expression for nonrelativistic quarkonium
spectrum:
\begin{equation}
\label{spectr}M_{nL}=m_0+m_1\left( n-\frac 12+\sqrt{\left(\nu+\frac 12\right)^2+k^2L(L+1)}%
\,\right) ^2-\frac{m_2}{\left( n-\frac 12+\sqrt{\left(\nu+\frac
12\right)^2+k^2L(L+1)}\,\right) ^2}
\end{equation}
where parameters are defined as follows: $$
m_0=2m+2\pi^2a^3k^4\varepsilon _0,\;\;m_1=\frac
1{ma^2k^4},\;\;m_2=\frac m4g^4k^4. $$ Energy, expended to
reconstruction of nonperturbative vacuum condensate inside
quarkonium, here is presented as $E_{vac}=2\pi ^2a^3k^4\varepsilon
_0$, where $V=2\pi ^2a^3k^4$ is quarkonium volume, calculated
under the metric of effective Riemannian space; $\varepsilon _0$ -
parameter quantitatively describing vacuum reconstruction. \

\section{Fine and hyperfine splitting.}

For the analysis of the relativistic effects it is necessary
beforehand to agree about interpretation of nonrelativistic levels
(\ref{spectr}). There are two possible interpretation. According
to the first of them the eigenvalues of the nonrelativistic
Hamiltonian are centers of gravity of the spin triplets i.e.
ortho$(S=1)$quarkonium levels, according to the second the
nonrelativistic Hamiltonian gives the centers of gravity of the
full spin multiplets. For the states with $L\neq 0$, this
interpretations are equivalent and the difference occurs only for
the $S$--states. In the framework of potential ideology there is
not justification for one interpretation rather than the other.
From the practical reasons that the masses of the spin singlets
for bottomonium are not known we interpret nonrelativistic levels
as the centers of gravity of orthoquarkonium.

In present work we shall be limited by consideration only
spin-dependent interactions -- $H_{SD}$. Following to standard
procedure, based on Breit-Fermi Hamiltonian \cite{Lau},
\cite{Lucha} and in conformity with general principles of
geometrized model, explained in Section III, let's write
$H_{SD}$ in the covariant form:%
\begin{equation}
\label{sd}H_{SD}=-\frac{3i}{2m^2}\varepsilon _{\alpha \beta \gamma }(s_Q^\alpha +s_{%
\overline{Q}}^\alpha )\left( \nabla ^\beta U\right) \left( \nabla
^\gamma \right)-\frac 1{m^2}(s_Q^\alpha \nabla _\alpha
)(s_{\overline{Q}}^\beta \nabla _\beta )U+\frac 1{m^2}\left[
\nabla ^\alpha \nabla _\alpha U\right] \left( \gamma _{\alpha
\beta }s_Q^\alpha s_{\overline{Q}}^\beta \right) ,
\end{equation}
where $\nabla _\alpha $ is the covariant derivation operator; $s_Q^\alpha
=\frac 12\sigma _Q^\alpha $ is (anti)quark spin operator. The matrixes $%
\sigma _Q^\alpha $ are defined in the effective Riemannian space by
condition
\begin{equation}
\label{spatial}\sigma _Q^\alpha \sigma _Q^\beta +\sigma _Q^\beta \sigma
_Q^\alpha =2\gamma ^{\alpha \beta }
\end{equation}
and covariant derivative\footnote{%
Expressions (\ref{spatial}) and (\ref{spatial1}) are
nonrelativistic limit of fundamental spatial-spin connection
$\gamma ^i\gamma ^k+\gamma ^k\gamma ^i=2g^{ik}$ and $\nabla
_i\gamma _k=0$.}
\begin{equation}
\label{spatial1}\nabla _\beta \sigma _Q^\alpha =0
\end{equation}
Representing the metric (\ref{3}) as follows
$$
dl^2=f^2(x,y,z)\left[ dx^2+dy^2+dz^2\right] =\gamma _{\alpha \beta
}dx^\alpha dx^\beta
$$
and further taking into account the spherical symmetry of potential and
ratio (\ref{uniq}) Hamiltonian (\ref{sd}) is reduced to
\begin{equation}
\label{sd1}H_{SD}=H_{LS}+H_T+H_{hyp},
\end{equation}
where
\begin{equation}
\label{ls1}H_{LS}=\frac 1{2m^2}\left( \frac{3g^2}{r^3f}\right) L_\alpha
S^\alpha ,
\end{equation}
\begin{equation}
\label{tensor}H_T=\frac 1{2m^2}\left( \frac{3g^2}{r^3f}\right) \left[
1+\frac 13\frac rf\frac{df}{dr}\right] \left\{ \frac{(S^\alpha x_\alpha
)(S^\beta x_\beta )}{r^2}-\frac 13S^\alpha S_\alpha \right\} ,
\end{equation}
\begin{equation}
\label{hyp}H_{hyp}=\frac 2{3m^2}\left[ \nabla _\beta \nabla ^\beta U\right]
\left( \frac{S^\alpha S_\alpha }2-\frac 34-\frac 14\right) =\frac{2g^2}{3m^2}%
\delta (\vec r)\left( \frac{S^\alpha S_\alpha }2-1\right) ,
\end{equation}
here $S^\alpha =s_Q^\alpha +s_{\overline{Q}}^\alpha $. $H_{LS}$ (\ref{ls1})
and $H_T$ (\ref{tensor}) are accordingly the spin-orbit and the tensor
interactions, responsible for the fine structure of orthoquarkonium levels, $%
H_{hyp}$ (\ref{hyp}) is the hyperfine interaction, splitting the
states ortho and para$(S=0)$quarkonium. The number $-\frac 14$ in
(\ref{hyp}) corresponds to our assumption that the nonrelativistic
Hamiltonian gives the orthoquarkonium levels.

The numerical values for the fine and hyperfine splitting of nonrelativistic
levels are found by calculation of matrix elements for the operators $H_{LS},$ $%
H_T,$ $H_{hyp}$ on wave functions of corresponding states and by
virtue of factorization (\ref{factorization}) the determination of
the matrix elements of spin and spatial operators can be carried
out separately.

Let us to consider at first the fine splitting. The matrix elements for the
spin operators in (\ref{ls1}) and (\ref{tensor}) are
\begin{equation}
\label{matrix ls}K_{JL}^{(LS)}=\left\langle J,L{\bf ,}1\left| {\bf LS}%
\right| J,L{\bf ,}1\right\rangle =\frac 12\left[ {\bf J}^2-{\bf L}%
^2-2\right]
\end{equation}
\begin{equation}
\label{matrix t}K_{JL}^{(T)}=\left\langle J,L{\bf ,}1\left| \left( n_\alpha
n_\beta -\frac 13\gamma _{\alpha \beta }\right) S^\alpha S^\beta \right| J,L%
{\bf ,}1\right\rangle =-\frac 13\left[ \frac{6\left( {\bf LS}\right)
^2+3\left( {\bf LS}\right) -4{\bf L}^2}{4{\bf L}^2-3}\right]
\end{equation}
where $n_\alpha =x_\alpha /r,\quad {\bf J=L+S}$. After transition in (\ref
{ls1}) and (\ref{tensor}) to coordinate $\eta $, we can write the final
expressions for numerical corrections as follows
\begin{equation}
\label{Els}\delta E_{nLJ}^{(LS)}=A_{nL}K_{JL}^{(LS)},\quad A_{nL}=\frac{3g^2k%
}{8m^2a^3}\int\limits_0^\pi \frac 1{\sin \eta \tan ^{2k}\frac \eta 2}\phi
_{nL}^2d\eta
\end{equation}
\begin{equation}
\label{Tls}\delta E_{nLJ}^{(T)}=B_{nL}K_{JL}^{(T)},\quad B_{nL}=\frac{3g^2k}{%
8m^2a^3}\int\limits_0^\pi \frac 1{\sin \eta \tan ^{2k}\frac \eta 2}\left(
\frac{2k+\cos \eta }{3k}\right) \phi _{nL}^2d\eta
\end{equation}
Let us to note that expressions (\ref{Els}) and (\ref{Tls}) do not contain
any free parameters.

When analyzing hyperfine interaction, we suppose additional
quark-condensate correlation, conditioned by influence of
spin-spin states of quarks on the state of nonperturbative vacuum
inside quarkonium. We understand that is reflected in the form of
interquark potential, formed with active participation of this
vacuum. For account with supposition we replace phenomenological
parameter $k$ (appropriating to orthoquarkonium) in the potential
(\ref{potential}) by a new phenomenological parameter $k_0$
appropriating to paraquarkonium, which value is fixed from the
adoption of the experimental data. Since there is not experimental
data for parabottomonium we shall not consider the spin-spin
interactions in this quarkonium.

Further we shall notes, that the $\delta$ function in (\ref{hyp})
comes from the unnatural nonrelativistic reduction and will become
a smooth function with a finite range if this is calculated
correctly. The standard way to calculate $H_{hyp}$ is to replace
$\delta (\vec r)$ by a smeared function \cite{Hyper}, which we
assume
$$
S(r)=\left[ \frac 1{4\pi f^3r^2}\right] \frac{\sin \left\{ \mu r\right\} }%
r,\quad \lim \limits_{\mu \rightarrow \infty }S(r)=\delta (\vec r)
$$ where $\mu $ is the range of smearing, which is a free
phenomenological parameter, we choose
$\mu=\frac{1}{3}\frac{1}{2a}.$

In terms of coordinate $\eta $ the numerical values of hyperfine corrections
are
\begin{equation}
\label{Lasthyp}\delta E_{nL}^{(hyp)}=-\frac{2g^2}{3m^2a^3k}\int\limits_0^\pi
\frac{\sin \left\{ \frac 13 \tan ^k\left( \frac \eta 2\right) \right\} }{\sin ^3\eta }%
\phi _{nL}^2d\eta ,\quad
\end{equation}
The criterion of adaptability of such mode calculation $H_{hyp}$ can be
following. For a P state, the center of gravity (COG) of triplet 1$^3$P$_{%
{\rm (COG)}}$ must coincide with singlet 1$^1$P$_1$ if $H_{hyp}$ is
pointlike interaction, since $\left| \Psi (0)\right| ^2=0$. From
experimental data for charmonium, one finds
$$
\quad 1^3{\rm P_{(COG)}-1^1P_1=1\ Mev }%
$$
Our calculation under the formula (\ref{Lasthyp}) give
$$
\quad 1^3{\rm P_{(COG)}-1^1P_1=5\ Mev }%
$$
This is a reasonable accuracy and this means the range of $H_{hyp}$ is
reasonable.

Finally, the paraquarkonium levels are determined in two stage: on the fist
stage the nonrelativistic levels (\ref{spectr}) with parameter $k_0$ are
found and on the second the corrections (\ref{Lasthyp}) are calculated.

Wave functions $\phi _{nL}$ used in expressions (\ref {Els}),
(\ref{Tls}) and (\ref {Lasthyp}) in correspondence with
(\ref{exact}) have a
following form%
$$ \phi _{1L}=C_{1L}\exp \left( -\frac{\gamma k^4\eta
}{l+1}\right) \sin ^{l+1}\eta ; $$ $$ \phi _{2L}=C_{2L}\frac{\exp
\left( -\frac{\gamma k^4\eta }{l+2}\right) \sin ^{l+2}\eta
}{(l+2)(l+1)\sin \eta }\left[ (2l+3)\gamma k^4\sin \eta
-(2l^3+9l^2+13l+6)\cos \eta \right] ; $$ $$
\phi_{3L}=C_{3L}\frac{\exp \left( -\frac{\gamma k^4\eta
}{l+3}\right) \sin ^{l+3}\eta }{(l+3)^2(l+1)\sin ^2\eta
}\biggl[(\gamma ^2k^8l+10\gamma ^2k^845-93l-65l^2-2l^4-19l^3)\sin
^2\eta $$ $$ +(4l^5+44l^4+187l^3-369l+381l^2+135)\cos ^2\eta
-(8\gamma k^4l^3+56\gamma k^4l^2+126\gamma k^4+90\gamma k^4)\cos
\eta \sin \eta \biggr ]; $$ and are normalized according to
(\ref{normirovka}).

\section{Results and conclusion.\ }

In Table 1 the values of parameters, appearing in expression for
nonrelativistic spectra (\ref{spectr}) are presented, which were
used at all stages of charmonium and bottomonium spectra numerical
modeling. In Table 2 the values of quarkonia physical parameters
are presented. Table 3 contain outcomes of charmonium and
bottomonium spectra modeling for experimentally observable levels.
On Fig. 1, 2 in standard representation the complete charmonium
and bottomonium spectra, including experimentally known levels and
our theoretical prediction for not yet opened levels are
presented. The levels have spectroscopic designation $^{2S+1}L_J$
and are located in corresponding to experimentally fixed quantum
numbers $J^{PC}$.

Accuracy of spectra theoretical modeling was estimated under the
formula
\begin{equation}
\label{ac}\delta =\sqrt{\frac 1N\sum_{i=1}^N\left( \frac{M_{i(th)}-M_{i(ex)}%
}{M_{i(ex)}-M_{1S}}\right) ^2}
\end{equation}
where $N$ - number of quarkonium levels, on which is conducted comparison
theories $M_{i(th)}$ and experiment $M_{i(ex)}$ behind elimination of ground
state of orthocharmonium $M_{1{\rm S}}=3097\;$MeV and orthobottomonium $M_{1%
{\rm S}}=9460\;$MeV, accepted for a count beginning. For all experimentally
known levels of charmonium $N=11$; in this case the calculation under the formula (%
\ref{ac}) give $\delta _{11}^{(c)}=4.06\cdot 10^{-2}$. Levels below charm
threshold is estimated under the same formula (\ref{ac}) with $N=7$; here we
have $\delta _7^{(c)}=5.05\cdot 10^{-2}$. For all experimentally known
levels of bottomonium $N=11$, here $\delta _{11}^{(b)}=3.28\cdot 10^{-2}$.
For levels below bottom threshold $N=8$, $\delta _8^{(b)}=0.96\cdot 10^{-2}$%
. In Table 4 the accuracy of spectra modeling of the several best models
known for us, calculated under the formula (\ref{ac}), are presented.

From numerical values of charmonium and bottomonium parameters,
presented in Tables 1 and 2, follows the quarkonia are essentially
nonperturbative objects. Foundations for this conclusion serve, at
first large value of a coupling constant $g^2$ on quarkonium
scale; secondly, essential difference of parameter $k$ from one,
showing the absence of Coulomb's asymptotics with $r\to 0$, that,
in turn, means essential nonperturbative properties of vacuum
inside quarkonium on all scales. The last conclusion is confirmed
by comparison of parameter $\varepsilon _0$, describing a power of
vacuum condensate partial reconstruction inside quarkonium, with
the module of outside-vacuum condensate energy density
\cite{vacuum}. $$ |\varepsilon _{out}|=0.3\cdot \left\langle
0\left| \frac \alpha \pi G_{\mu \nu }^aG^{a\mu \nu }\right|
0\right\rangle \approx (275MeV)^4 $$ As we see, the deconfinement
of quarks within quarkonium take place already with $\varepsilon
_0/|\varepsilon _{out}|\approx 0.03$ for charmonium and with
$\varepsilon _0/|\varepsilon _{out}|\approx 0.05$ for bottomonium.
Conclusion about more force reconstruction of nonperturbative
vacuum inside bottomonium in comparison with charmonium is
qualitatively coordinated with values of nonperturbative coupling
constant $g^2$: in bottomonium it is less, than in charmonium.

In the end let us to make conclusion: in the framework of our
model we can quantitatively see, that nonperturbative vacuum play
key role in hadron structure, it is necessary and defining
ingredient of quarks dynamics and, therefore observable properties
of quarkonia. By justification of this conclusion the good fit to
experimental data serve.
\section{Acknowledgments}
I am very grateful to G.M. Vereshkov for discussions and valuable
advises.
\newpage
\begin{table}[p]\centering
\caption{ Parameters of model.}
\begin{tabular}{|c|c|c|}
\hline
      {\bf Parameter}      &  {\bf Charmonium} & {\bf Bottomonium}  \\ \hline
   $ \nu $                & $0.158  $        &   $0.139  $        \\
   $ 2k^2_0 $             & $1.291   $        &   -- -- --        \\
   $ 2k^2 $             & $1.179   $        & $1.360  $          \\
   $ m_0 $                & $3628.209 \;$MeV & $9963.116 \;$MeV   \\
   $ m_1 $                & $49.348 \;$Mev   & $48.136 \;$Mev     \\
   $ m_2 $                & $ 800.825 \;$Mev & $ 734.050 \;$Mev   \\ \hline
\end{tabular}
\end{table}

\begin{table}[p]\centering
\caption{ Parameters of quarkonia.}
\begin{tabular}{|c|c|c|}
\hline
      {\bf Parameter}      &  {\bf Charmonium}       & {\bf Bottomonium}      \\ \hline
     $m $            & $ 1700 \;$MeV             & $ 4905 \;$MeV        \\
     $ g^2 $            & $ 2.328$               & $ 1.138$               \\
     $ a $                & $ 5.856 \cdot10^{-3}\; $MeV$^{-1}$& $ 3.028\cdot10^{-3}\; $MeV$^{-1}$\\
     $ \varepsilon_0 $    & $ (113.442 \;$MeV$)^4$   & $ (129.298 \;$MeV$)^4$   \\
     $ \Lambda$          &  $57.243 \;$MeV         & $68.278 \;$MeV          \\ \hline
\end{tabular}
\end{table}
\begin{table}[p]\centering
\caption{\ Fitted levels of quarkonia (MeV).}
\begin{tabular}{|c|c|c||c|c|c|}
\hline \multicolumn{3}{|c||}{\bf
Charmonium}&\multicolumn{3}{c|}{\bf Bottomonium}\\ \hline {\bf
State}&{Data\cite{data}}&{This work}&{\bf
State}&{Data\cite{data}}&{This work }\\ \hline
$1^1$S$_0$&2980&2974&$1^3$S$_1$&9460&9460\\ $1^3$S$_1$&3097&3097
&$1^3$P$_0$&9860 &9860\\ $1^1$P$_1$&3526&3512&$1^3$P$_1$&9892
&9901\\ $1^3$P$_0$&$3415$&3413&$1^3$P$_2$&9913&9916\\
$1^3$P$_1$&3511&3517&$2^3$S$_1$&10023&10023\\
$1^3$P$_2$&3556&3556&$2^3$P$_0$&10232&10222\\
$2^1$S$_0$&3594&3647&$2^1$P$_1$&10255&10254\\
$2^3$S$_1$&3686&3686&$2^3$P$_2$&10269&10266\\
$1^3$D$_1$&$3770$&3772&$3^1$S$_1$&10355&10363\\
$3^3$S$_1$&$4040$&4040&$2^3$D$_1$ &10580& 10522\\
$2^3$D$_1$&$4159\pm 20$&4127&$4^3$S$_1$ &10865 &10745 \\
$4^3$S$_1$&$4415\pm 6$&4435&$3^3$D$_1$ &11020&10936\\ \hline
\end{tabular}
\end{table}
\begin{table}[p]\centering
\caption{ Accuracy of spectra modeling (\%).}
\begin{tabular}{|c|c|c|c|c|c|c|}
\hline {\bf Accuracy}&{\bf EQ}\cite{Eichten}&{\bf
GJ}\cite{Gupta}&{\bf MZ}\cite{Zalewski} &{\bf ZOR}\cite{Zeng}&{\bf
BBZ} \cite{Baker}&{ This work}\\ \hline $\delta _7^{(c)}$ &6.36
&1.9 &2.37 &9.83&5.34 &5.05  \\ $\delta _{11}^{(c)}$  &--- &--- &
--- &9.83&--- &4.06
\\ \hline $\delta _8^{(b)}$    &5.15 &--- &0.14 &3.25&2.04 &0.96 \\
$\delta _{11}^{(b)}$  &---     &--- &
---    & 4.85&--- &3.28\\ \hline
\end{tabular}
\end{table}
{%
\begin{figure}[p]
\unitlength 0.47mm \linethickness{0.4pt}{\sc
\begin{picture}(0,210)
\tiny \put(20,0){\framebox(300,210)}
\put(20,10){\line(1,0){2}}\put(2,6){\normalsize2.80}
\put(20,50){\line(1,0){2}}\put(2,46){\normalsize3.20}
\put(20,90){\line(1,0){2}}\put(2,86){\normalsize3.60}
\put(20,130){\line(1,0){2}}\put(2,126){\normalsize4.00}
\put(20,170){\line(1,0){2}}\put(2,166){\normalsize4.40}
\put(2,206){\normalsize4.80}
\put(45,0){\line(0,1){2}}
\put(70,0){\line(0,1){2}}
\put(95,0){\line(0,1){2}}
\put(120,0){\line(0,1){2}}
\put(145,0){\line(0,1){2}}
\put(170,0){\line(0,1){2}}
\put(195,0){\line(0,1){2}}
\put(220,0){\line(0,1){2}}
\put(245,0){\line(0,1){2}}
\put(270,0){\line(0,1){2}}
\put(295,0){\line(0,1){2}}

\put(26,-8){\normalsize$0^{-+}$}
\put(51,-8){\normalsize$1^{--}$}
\put(76,-8){\normalsize$1^{+-}$}
\put(101,-8){\normalsize$0^{++}$}
\put(126,-8){\normalsize$1^{++}$}
\put(151,-8){\normalsize$2^{++}$}
\put(176,-8){\normalsize$2^{-+}$}
\put(201,-8){\normalsize$2^{--}$}
\put(226,-8){\normalsize$3^{--}$}
\put(251,-8){\normalsize$3^{+-}$}
\put(276,-8){\normalsize$3^{++}$}
\put(301,-8){\normalsize$4^{++}$}
\put(21,27){\line(1,0){12}}\put(20.5,22){$1^1$S$_0(2.97)$}
\put(21,95){\line(1,0){24}}\put(20.5,96){$2^1$S$_0(3.65)$}%
\put(21,128){\line(1,0){24}}\put(20.5,130){$3^1$S$_0(3.98)$}%

\put(46,40){\line(1,0){12}}\put(46,41.7){$1^3$S$_1(3.10)$}
\put(46,99){\line(1,0){12}}\put(48,93.7){$2^3$S$_1(3.69)$}
\put(46,134){\line(1,0){12}}\put(48,128.7){$3^3$S$_1(4.04)$}
\put(46,174){\line(1,0){12}}\put(46,176){$4^3$S$_1(4.44)$}

\put(71,81){\line(1,0){24}}\put(68,76.3){$1^1$P$_1(3.51)$}
\put(71,117){\line(1,0){24}}\put(68,119){$2^1$P$_1(3.87)$}
\put(71,150){\line(1,0){24}}\put(68,152){$3^1$P$_1(4.20)$}

\put(96,71){\line(1,0){12}}\put(96,65.7){$1^3$P$_0(3.41)$}
\put(121,83){\line(1,0){24}}\put(118.5,85){$1^3$P$_1(3.52)$}
\put(146,86){\line(1,0){12}}\put(146,88){$1^3$P$_2(3.56)$}

\put(96,112){\line(1,0){24}}\put(96,114){$2^3$P$_0(3.82)$}
\put(121,119){\line(1,0){24}}\put(117,121){$2^3$P$_1(3.90)$}
\put(146,121){\line(1,0){24}}\put(146,123){$2^3$P$_2(3.92)$}

\put(96,148){\line(1,0){24}}\put(96,150){$3^3$P$_0(4.18)$}
\put(121,157){\line(1,0){24}}\put(118,159){$3^3$P$_1(4.27)$}
\put(146,160){\line(1,0){24}}\put(146,162){$3^3$P$_2(4.33)$}
\put(171,110){\line(1,0){24}}\put(168,112){$1^1$D$_2(3.80)$}
\put(171,142){\line(1,0){24}}\put(169.5,144){$2^1$D$_2(4.12)$}%
\put(171,178){\line(1,0){24}}\put(169.5,180){$3^1$D$_2(4.48)$}%

\put(46,107){\line(1,0){12}}\put(46,109){$1^3$D$_1(3.77)$}
\put(196,110){\line(1,0){24}} \put(194,112){$1^3$D$_2(3.80)$}
\put(221,116){\line(1,0){24}} \put(221,118){$1^3$D$_3(3.86)$}

\put(46,143){\line(1,0){24}} \put(46,137.5){$2^3$D$_1(4.13)$}
\put(196,146){\line(1,0){24}} \put(196,141){$2^3$D$_2(4.16)$}
\put(221,148){\line(1,0){24}} \put(221,150){$2^3$D$_3(4.18)$}

\put(46,184){\line(1,0){24}} \put(46,186){$3^3$D$_1(4.54)$}
\put(196,198){\line(1,0){24}} \put(194,200){$3^3$D$_2(4.58)$}
\put(221,201){\line(1,0){24}} \put(221,203){$3^3$D$_3(4.61)$}

\put(246,135){\line(1,0){24}}\put(242,137){$1^1$F$_3(4.05)$}

\put(146,135){\line(1,0){24}}\put(142,137){$1^3$F$_2(4.05)$}
\put(271,137){\line(1,0){24}}\put(267,139){$1^3$F$_3(4.07)$}
\put(296,139){\line(1,0){24}}\put(293,141){$1^3$F$_4(4.09)$}

\put(20,101){\dashbox{3}(300,0)}\put(250,103){ \bf \footnotesize
Charm threshold}
\put(250,0){\framebox(70,30)}%
\put(272,20){\footnotesize data}%
\put(272,10){\footnotesize theory}%
\multiput(254,22)(1.2,0){14}{.}%
\put(254,12){\line(1,0){16}}%
\multiput(58,39.7)(1.2,0){10}{.}
\multiput(32.5,27.7)(1.2,0){10}{.}
\multiput(20,88.7)(1.2,0){20}{.}
\multiput(58,98.7)(1.2,0){10}{.}
\multiput(58,133.7)(1.2,0){10}{.}
\multiput(58,171.7)(1.2,0){10}{.}
\multiput(71,82.7)(1.2,0){20}{.} \multiput(108,71.7)(1.2,0){10}{.}
\multiput(121,80.7)(1.2,0){20}{.}
\multiput(158,85.7)(1.2,0){10}{.}
\multiput(57.5,106.7)(1.2,0){10}{.}
\multiput(46,146)(1.2,0){20}{.}
\end{picture}
} \put(0,-16){\caption{Charmonium spectrum (GeV).}}
\end{figure}
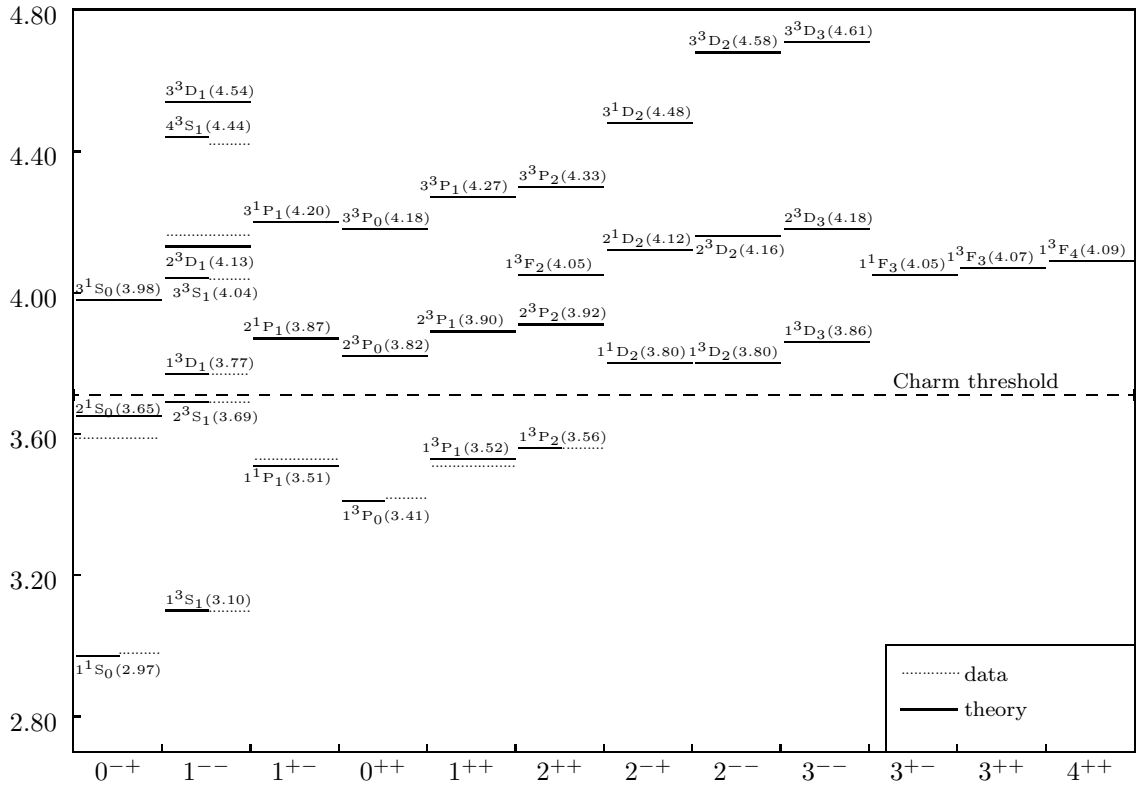}
{%
\begin{figure}[p]
\unitlength 0.47mm \linethickness{0.4pt}{\sc
\begin{picture}(0,230)
\tiny
\put(20,0){\framebox(300,230)}
\put(20,10){\line(1,0){2}}\put(2,6){\normalsize9.20}
\put(20,50){\line(1,0){2}}\put(2,46){\normalsize9.60}
\put(20,90){\line(1,0){2}}\put(-2,86){\normalsize10.00}
\put(20,130){\line(1,0){2}}\put(-2,126){\normalsize10.40}
\put(20,170){\line(1,0){2}}\put(-2,166){\normalsize10.80}
\put(20,210){\line(1,0){2}}\put(-2,206){\normalsize11.20}
\put(45,0){\line(0,1){2}}
\put(70,0){\line(0,1){2}}
\put(95,0){\line(0,1){2}}
\put(120,0){\line(0,1){2}}
\put(145,0){\line(0,1){2}}
\put(170,0){\line(0,1){2}}
\put(195,0){\line(0,1){2}}
\put(220,0){\line(0,1){2}}
\put(245.5,0){\line(0,1){2}}
\put(270,0){\line(0,1){2}}
\put(295,0){\line(0,1){2}}

\put(26,-8){\normalsize$0^{-+}$}
\put(50,-8){\normalsize$1^{--}$}
\put(76,-8){\normalsize$1^{+-}$}
\put(100,-8){\normalsize$0^{++}$}
\put(124,-8){\normalsize$1^{++}$}
\put(148,-8){\normalsize$2^{++}$}
\put(176,-8){\normalsize$2^{-+}$}
\put(200,-8){\normalsize$2^{--}$}
\put(225,-8){\normalsize$3^{--}$}
\put(250,-8){\normalsize$3^{+-}$}
\put(275,-8){\normalsize$3^{++}$}
\put(300,-8){\normalsize$4^{++}$}
\put(46,36){\line(1,0){12}}\put(46,38){$1^3$S$_1(9.46)$}
\put(46,92){\line(1,0){12}}\put(46,94){$2^3$S$_1(10.02)$}
\put(46,126){\line(1,0){12}}\put(46,128){$3^3$S$_1(10.36)$}
\put(46,165){\line(1,0){24}}\put(46,167){$4^3$S$_1(10.75)$}
\put(46,211){\line(1,0){24}}\put(46,213){$5^3$S$_1(11.21)$}

\put(96,76){\line(1,0){12}}\put(92,78){$1^3$P$_0(9.86)$}
\put(121,80){\line(1,0){12}}\put(119,82){$1^3$P$_1(9.90)$}
\put(146,82){\line(1,0){12}}\put(146,84){$1^3$P$_2(9.92)$}
\put(96,112){\line(1,0){12}}\put(91,106.5){$2^3$P$_0(10.22)$}
\put(121,115.5){\line(1,0){12}}\put(121,109.5){$2^3$P$_1(10.25)$}
\put(146,118){\line(1,0){12}}\put(146,120){$2^3$P$_2(10.27)$}
\put(96,148){\line(1,0){24}}\put(86,150){$3^3$P$_0(10.58)$}
\put(121,152){\line(1,0){24}}\put(116,154){$3^3$P$_1(10.62)$}
\put(146,153){\line(1,0){24}}\put(146,155){$3^3$P$_2(11.63)$}

\put(46,107){\line(1,0){24}}\put(44,109){$1^3$D$_1(10.17)$}
\put(196,109){\line(1,0){24}}\put(191,111){$1^3$D$_2(10.19)$}
\put(221,111){\line(1,0){24}}\put(223,113){$1^3$D$_3(10.21)$}
\put(46,142){\line(1,0){24}}\put(40,144){$2^3$D$_1(10.52)$}
\put(196,144){\line(1,0){24}}\put(191,146){$2^3$D$_2(10.54)$}
\put(221,146){\line(1,0){24}}\put(221.5,148){$2^3$D$_3(10.56)$}
\put(46,184){\line(1,0){24}}\put(46,186){$3^3$D$_1(10.94)$}
\put(196,185){\line(1,0){24}}\put(190,187){$3^3$D$_2(10.95)$}
\put(221,186){\line(1,0){24}}\put(221,188){$3^3$D$_3(11.96)$}

\put(146,136){\line(1,0){24}}\put(146,130.5){$1^3$F$_2(10.46)$}
\put(271,137){\line(1,0){24}}\put(259,132){$1^3$F$_3(10.47)$}
\put(296,137){\line(1,0){24}}\put(290,132){$1^3$F$_4(10.47)$}
\put(20,140){\dashbox{4}(300,0)} \put(260,142){\bf \footnotesize
Bottom threshold  }
\put(250,0){\framebox(70,30)}%
\put(272,20){\footnotesize data}%
\put(272,10){\footnotesize theory}%
\multiput(254,22)(1.2,0){14}{.}%
\put(254,12){\line(1,0){16}}%
\multiput(58,35.8)(1.2,0){10}{.}
\multiput(58,91.7)(1.2,0){10}{.}
\multiput(58,125.7)(1.2,0){10}{.}
\multiput(46,175.9)(1.2,0){20}{.}
\multiput(46,192)(1.2,0){20}{.}
\multiput(108,75.7)(1.2,0){10}{.}
\multiput(133,78.7)(1.2,0){10}{.}
\multiput(158,80.7)(1.2,0){10}{.}
\multiput(108,112.9)(1.2,0){10}{.}
\multiput(133,115.6)(1.2,0){10}{.}
\multiput(158,117.7)(1.2,0){10}{.}
\multiput(46,147.9)(1.2,0){20}{.}
\end{picture}
} \put(0,-16){\caption{Bottomonium spectrum (GeV).}}
\end{figure}
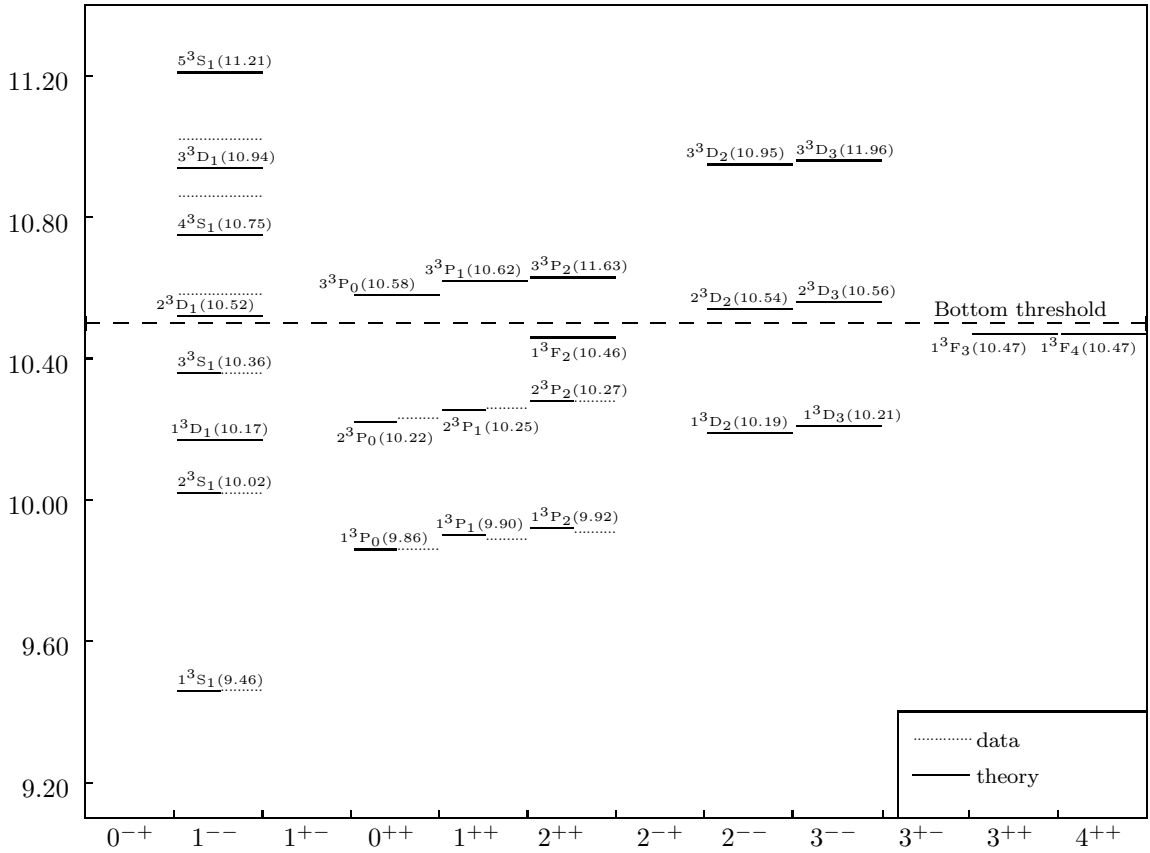}

\end{document}